\documentclass[prd,aps]{revtex4}

\begin{document}
\preprint{{hep-th/0502230} \hfill {UCVFC-DF-17-2005}}
\title{Loop Equations in Abelian Gauge Theories}
\author{Cayetano Di Bartolo$^{1}$,
Lorenzo Leal$^{2}$ and Francisco Pe\~{n}a$^{1,2}$}

\affiliation { 1. Departamento de F\'{\i}sica, Universidad Sim\'on
Bol\'{\i}var,\\ Aptdo. 89000, Caracas 1080-A, Venezuela.\\ 2.
Grupo de Campos y Part\'{\i}culas, Departamento de F\'{\i}sica,
Facultad de Ciencias, Universidad Central de Venezuela, AP 47270,
Caracas 1041-A, Venezuela. }

\begin{abstract}
The equations obeyed by the vacuum expectation value of the Wilson
loop of Abelian gauge theories are considered from the point of
view of the loop-space. An  approximative scheme  for studying
these  loop-equations for lattice Maxwell theory is presented. The
approximation leads to a partial difference equation in the area
and length variables of the loop, and certain physically motivated
ansatz is seen to reproduce the mean field results from a
geometrical perspective.
\end{abstract}

\maketitle

\section{Introduction}

The loop representation \cite{uno,dos,tres,cuatro,cinco} is a
useful tool for studying non-perturbative features of gauge
theories, both in the lattice and the continuous frameworks.
Loop-space formulations of quantum gravity have also been
developed, giving rise to a geometrical setting in which some of
the long-standing questions concerning the small distance behavior
of space-time can be properly addressed
\cite{Gra1,Gra2,Gra3,Gra4,Gra5}.

Some years ago, the loop representation formulation of Hamiltonian
lattice gauge theories was studied for several models
\cite{lat1,lat2,lat3,lat4}. The goal of that formulation was to
produce  loop-dependent Schr\"odinger equations, and to develop
loop-based schemes of approximation to study the spectra and the
phase-transitions of the theories. The purpose of this paper is to
draw some preliminary lines for a similar study within the
Lagrangian formulation. Concretely, we shall study  the
loop-equation that obeys the Wilson loop average $\langle
W(C)\rangle$ of Maxwell theory in the lattice. We shall be
interested in exploring some basic aspects of the loop content of
the theory rather than in obtaining accurate predictions for the
relevant observables.

As a warm-up for the lattice study, we first recall briefly some
basic facts about  loop-equations for Abelian gauge theories in
the continuum. As it is well known, the loop-equation
\cite{Poly,MM1,MM2,M3} (sometimes called Migdal, Polyakov or
Schwinger-Dyson equation) for $\langle W(C)\rangle$ can be exactly
solved for the free Maxwell theory within the continuous framework
\cite{mig}.

In the non-Abelian cases, loop-equations have been considered
mainly within the large $N$ approximation
\cite{thooft,MM1,MM2,M3}. Some recent applications of the
loop-equation approach to QCD can be found in references
\cite{jugeau,baldicchi,brambilla}.  On the other hand, in the
lattice, even the Abelian cases are non-trivial (except in two
dimensions), due to the presence of the phase transition that
separates the confining regime (absent in the formulation in the
continuum)  from the weak coupling one.

The paper is organized as follows. In section two we review the
Wilson loop equations for Maxwell and Chern-Simons theories in the
continuum from a loop-space perspective. In the last section we
consider Maxwell theory in the lattice.

\section {Maxwell and Chern-Simons Wilson-loop equations in the continuum}

We shall deal with the vacuum expectation value
\begin{equation} \label{wilson1}
\langle W(C)\rangle \equiv \int \textit{D}A \,\,W(C) \,\,\exp iS,
\end{equation}
of the Wilson loop
\begin{equation} \label{wilson loop}
W(C) \equiv \exp (-ie\oint_{C} dx^{\mu} A_{\mu}(x)),
\end{equation}
where $S$ is the action functional of an Abelian gauge theory,
that we shall first take as the n-dimensional Maxwell one
\begin{equation} \label{Max}
S_{Maxwell}= -\frac{1}{4}\int\, d^{n}x \,F_{\mu \nu}\,F^{\mu \nu}.
\end{equation}

Since $S_{Maxwell}$ is quadratic in the fields, the functional
integration in equation (\ref{wilson1}) can be performed. Instead,
we are interested in studying the functional differential equation
that $\langle W(C)\rangle $ obeys, i.e., in the Migdal
loop-equation of the model. To this end, we shall use a  few tools
of the loop-space formulation of gauge theories
\cite{uno,dos,tres,cuatro,cinco,C}.

Consider the space of oriented and piecewise continuous curves in
$R^{n}$.  We say that  curves $\gamma$ and $\gamma'$ are
equivalent if they share the same form factor $T^{\mu}(x,\gamma)$,
defined as
\begin{equation}\label{3.11}
T^{\mu}(x,\gamma)\equiv \int_{\gamma}
dy^{\mu}\,\delta^{(n)}(\vec{x}-\vec{y}).
\end{equation}
Every equivalence class defines a path. The composition of curves,
together with the equivalence relation stated above, defines a
group product among paths.  It can be shown that this group is
Abelian. Now, let us consider path-dependent functionals
$\Psi(\gamma)$. We introduce the path derivative $\delta_\mu (x)$,
that measures the change in $\Psi(\gamma)$ when an infinitesimal
path $u$ is attached to the argument $\gamma$ of $\Psi(\gamma)$ at
the point $x$, up to the first order in the vector $u^{\mu}$
associated to the path
\begin{equation}\label{3.12}
\Psi( u.\gamma ) = (1 + u^{\mu} \delta_{\mu} (x))\Psi (\gamma).
\end{equation}

We shall also use the loop derivative $\Delta_{\mu \nu}(x)$
defined as
\begin{equation}\label{3.13}
\Psi (\sigma \cdot \gamma)= \left(1+ \sigma^{\mu \nu}\Delta_{\mu
\nu}(x)\right)\, \Psi (\gamma),
\end{equation}
with   $\sigma^{\mu \nu}$ being the area enclosed by an
infinitesimal loop $\sigma$ attached at the point $x$. The loop
derivative is readily seen to be the curl of the path derivative
\begin{equation}\label{3.19} \Delta_{\mu \nu}(x)=
\partial_{\mu} \delta_{\nu}(x)-\partial_{\nu}
\delta_{\mu}(x).
\end{equation}
Also, we have
\begin{equation}\label{3.14a}
\delta_{\mu}(x)T^{\nu}(y,
\gamma)=\delta_{\mu}^{\nu}\,\delta^{(n)}(x-y),
\end{equation}
as can be readily shown.

Since $\langle W(C)\rangle$ is a genuine path-space function [it
depends on closed paths $C$], it makes sense to take its loop
derivative. Using the last equation, it is easy to see that

\begin{equation}\label{derivadaw}
\Delta_{\mu\nu}(x)\langle W(C)\rangle = -ie\int\textit{D}A\,
F_{\mu\nu}(x) W(C)\exp{\{iS\}}.
\end{equation}
Taking the divergence of this expression and recalling  the
variational principle for the Maxwell equations, we find
\begin{equation}\label{divergencia}
\partial^{\mu}\Delta_{\mu\nu}(x)\langle W(C)\rangle = e\int\textit{D}A\,\frac{\delta W(C)}{\delta
A^{\nu}(x)}e^{iS},
\end{equation}
where we have also integrated by parts in the functional sense.
Using
\begin{equation}\label{der-funcional}
\frac{\delta W(C)}{\delta A^{\nu}(x)} =
-ieg_{\mu\nu}T^\mu(x,C)W(C),
\end{equation}
we finally obtain the desired loop-equation
\begin{equation}\label{wl-maxwell}
\partial^{\mu}\Delta_{\mu\nu}\langle W(C)\rangle = -ie^2g_{\mu\nu}T^{\mu}(x,C)\langle W(C)\rangle.
\end{equation}

It should be stressed that everything in equation
(\ref{wl-maxwell}) is a true loop-dependent object, in the sense
of the equivalence classes mentioned at the beginning of this
section. Equation (\ref{wl-maxwell}) is a loop-space version of
the  ordinary differential equation
\begin{equation}\label{4.23}
a\frac{dF}{dx}(x)=xF(x),
\end{equation}
whose solution is $F(x)= \exp (\frac{x^{2}}{2a})$. Following this
 hint, one can readily obtain the solution
\begin{equation}\label{4.28}
\langle W(C)\rangle = K\exp{\left[\int d^nx\int
d^ny\,T^{\mu}(x,C)D_{\mu\nu}(x-y)T^{\nu}(y,C)\right]},
\end{equation}
where $D_{\mu\nu}(x-y)$ is the Feynman propagator of the theory.
Finally
\begin{equation}\label{4.63}
\langle W(C)\rangle = K\exp{\left[-\frac{e^2}{8\pi^2}\oint_c
dx^\mu\oint_c dy^\nu \frac{g_{\mu\nu}}{\mid x-y \mid^2}\right]}
\end{equation}
as it should be \cite{peskin}.

In the preceding discussion, we ignored that, due to gauge
invariance, the Wilson loop average given by equation
(\ref{wilson1}) is ill-defined. To remedy this, one may add the
gauge-fixing term $-\frac{1}{2\xi}(\partial_{\mu}A^{\mu})^{2}$ to
the Maxwell action (\ref{wilson loop}). In that case, equation
(\ref{wl-maxwell}) changes into
\begin{equation}\label{fad-popov}
\left[\partial^\mu\Delta_{\mu\nu}(x)
+\frac{1}{\xi}\partial_\nu\partial^\mu\delta_\mu(x)\right]\langle
W(C)\rangle = -ie^2g_{\mu\nu}T^\mu(x,C)\langle W(C)\rangle,
\end{equation}
but the net result (\ref{4.63}) does not change.

To conclude this section, let us briefly consider  the
Chern-Simons model, a topological theory whose action is
\begin{equation} \label{C-S}
S[A]=\int\mathrm d^3x \,
\varepsilon^{\mu\nu\lambda}A_{\mu}\partial_{\nu}A_{\lambda}.
\end{equation}
Now the differential equation for the Wilson loop is
\begin{equation} \label{4.114}
\varepsilon^{\mu\nu\lambda}\Delta_{\nu\lambda}(x)\langle
W(C)\rangle =-ie^2T^{\mu}(x,C)\langle W(C)\rangle.
\end{equation}
It is interesting to notice that, being the model topological, its
associated loop-equation is metric independent. This contrasts
with what occurs in Maxwell theory.

As before, the solution to the loop-equation should be an
exponential quadratic in the ``loop-coordinates" $T^{\mu}(x,C)$.
In fact, it is given by
\begin{equation} \label{4.136}
\langle W(C)\rangle = \exp{\left[-\frac{ie^2}{4}G(C,C)\right]},
\end{equation}
where
\begin{equation} \label{4.135}
G(C,C) = \frac{1}{4\pi}\oint_c dx^{\mu}\oint_c
dy^{\nu}\varepsilon_{\mu\nu\lambda}\frac{(x-y)^\lambda}{\mid x-y
\mid^3}
\end{equation}
is the Gauss self-linking number of the loop C, which is a well
known knot-invariant. Again, this result coincides with what is
obtained by performing the functional integral (\ref{wilson1}) for
the Chern-Simons model \cite{GaussNumber}.

This way of calculating $\langle W(C)\rangle$  can  also be used
for other Abelian gauge theories, such as the Maxwell-Chern-Simons
\cite{jackiw} and the Proca model in the Stueckelberg formulation
\cite{C}, since the differential loop-equations are similar to the
Maxwell loop-equation (\ref{wl-maxwell}).

\section {Maxwell Wilson-loop equation in the lattice}
Now we turn to consider loop-equations in the lattice. The
partition function of compact electrodynamics, which is the model
we are going to study, is given by
\begin{equation} \label{particion}
  Z=\int^{2\Pi}_{0} (\prod_{l>0} d\theta_l)\exp (-S),
\end{equation}
with
\begin{equation} \label{accion}
  S=-\beta \sum_{p>0}(W(p)+W^{\ast}(p)),
\end{equation}
and
\begin{equation} \label{plaqueta}
W(p)=\prod _{l\in p} \exp (i\theta _{l}).
\end{equation}
The Wilson functional $W(C)$\,\, is defined, as usual, by
\begin{equation}\
  W(C)=\prod_{l\epsilon C}\,exp \, (i\theta_l)\,\, .
\end{equation}
\noindent  In these expressions, $l$, $p$ and $C$ denote links,
plaquettes and closed loops respectively, $\beta$ is the inverse
of the tem\-pe\-ra\-tu\-re and $\theta_l$ is the angular variable
associated to link $l$.

The Schwinger-Dyson  equation for the Wilson loop, that
corresponds to the lattice version of equation (\ref{wl-maxwell}),
may be written as \cite{xue}
\begin{equation} \label{SDU1}
  \beta\,\sum_p \Delta(l,p)\,<W(p\cdot C)>+\Delta(l,C)\,<W(C)>=0,
\end{equation}
where
\begin{equation}
  \Delta(l,C)\equiv \sum_{l'\epsilon C}
  \widetilde{\delta}_{ll'}\quad,
\end{equation}
and
\begin{equation}\
  \widetilde{\delta}_{ll'}\equiv \left\{ \begin{array}{rl}
    1 & \mbox{for } l=l' \\
    -1 & \mbox{for } l=\bar{l}'  \\
    0 & \mbox{otherwise} .\
  \end{array} \right.
\end{equation}

\noindent  By $\bar{l}$ we mean the opposite of link $l$, and $<\,
>$ means statistical average. In equation (\ref{SDU1}) the sum
runs over all the plaquettes of the lattice, regardless of their
orientation. The factor $\Delta(l,p)$, however, restricts the sum
to the plaquettes attached to the link $l$. The loop product
$p\cdot C$ must be understood as the yuxtaposition of plaquette
$p$ and loop $C$, yielding a new closed loop. This corresponds to
the lattice version of the loop-derivative of
Gambini-Tr\'{i\'{}}as \cite{uno,cuatro} given in equation
(\ref{3.13}). Unlike non-Abelian loops, U(1) loops commute (i.e.,
they are not ``ordered") and do not have a marked starting point.

 Equation (\ref{SDU1}) arises by equating the averages of the Wilson loop
 for two different values of the phase factor $\exp i\theta _{l}$ assigned to a fixed
link $l$ of the lattice. This equality is  a consequence of the
invariance of the measure under $U(1)$ transformations. Expression
(\ref{SDU1}), which indeed represents one equation for each link
of the lattice, may be summed over all the links belonging to the
loop $C$ to produce the single equation
\begin{equation}\label{sumaSD}
  \beta\, \sum_{l\epsilon C} \sum_p \Delta(l,p) <\,W(p\cdot C)\,>
+  \Lambda_C<\,W(C)\,>=0,
\end{equation}
\noindent where
\begin{equation} \label{L2C}
  \Lambda_C \equiv \sum_{l\epsilon C}\sum_{l'\epsilon C}
\delta_{ll'}=\sum^{\infty}_{i=1} i^2L_i,
\end{equation}

\noindent with $L_i$ being the number of links that appear $i$
times in the loop $C$. For simple loops (i.e., loops without
multiple links) $\Lambda_C$ coincides with the length of $C$.

Equation  (\ref{sumaSD}) (or (\ref{SDU1})) is a  loop-equation,
since any reference to the electromagnetic potential  has been
summed when averaging over the different field configurations.
Despite its simple appearance (it is linear and first order in the
lattice ``loop derivative") and unlike its continuous counterpart,
it is not possible, as far as we know, to solve it exactly (except
for $D=2$). We shall discuss a simple method to extract some
information of this equation that relies on similar ideas that
were applied to the Hamiltonian lattice formulation in the past
\cite{lat1,lat3}.

To characterize completely a loop one needs an infinite amount of
variables. An incomplete list of them could contain the length,
area, number and types of corners, the class of knottiness to
which the loop belongs, and quite a few other variables. One could
select, on physical grounds, some of these variables and ignore
the (necessarily infinite number of) remaining ones. Another point
of view could be to  consider, instead, all the loops up to a
given size, which leads to a linked-cluster approach \cite{lat2}.
Here we shall adopt the first point of view, i.e., a ``collective
variables" approach, which has already been considered to deal
with the $Z_2$ Hamiltonian gauge theory \cite{lat1,lat3}. To
motivate the choice of variables we are going to make, we can
reason as follows.

It is well known that for ``big" loops and $D>3$, the Wilson
functional decreases exponentially with: a) the length, for the
weak coupling regime $(\beta =1/g^2\rightarrow\,+\infty)$\,; b)
the area, for the strong coupling one ($g$ is the coupling
constant, or the square root of the temperature)
\cite{bhanot,kogut,drou}. For $D=2$, the theory can be solved
exactly, and the area dependence holds for any value of the
coupling constant (i.e., the theory presents a single confining
phase). For $D=3$, approximation schemes and numerical simulations
yield that the situation seems to be very similar to the $D>3$
cases, although the exact solution has not been found
\cite{bhanot,kogut,drou}. All this suggests that length and area
are good candidates to describe, in a first approximation, the
Wilson loop functional. Moreover, the length variables\, $L_i$
already appear explicitly in the loop-equation (\ref{sumaSD}).
With this guide in mind, we approximate $<\,W(C)\,>$ in the form
\begin{equation}
  <\,W(C)\,>\approx f(L_1,L_2,\cdots ;A_1,A_2,\cdots )\quad ,
\end{equation}
\noindent where $A_i$ is the number of plaquettes of multiplicity
$i$ appearing in the loop $C$.

 This guess about the possible
relevant loop-variables should be accompanied by some restrictions
on the loops that we are going to take into account. Since we are
not including ``corners" in the list of variables, it seems
reasonable to ask our loops to be composed of lengthy straight
paths, which could be formed by multiple links. Furthermore, to
avoid ambiguities about how the first  term in equation
(\ref{sumaSD}) modifies the area variables, we must restrict
ourselves to consider planar loops.

We are ready to write down equation (\ref{sumaSD}) for
$f(L_i;A_i)$. The first term in this equation produces several
types of contributions, depending on how the plaquette $p$ is
appended to the loop $C$. Let us begin with

\begin{equation}
  +(2D-3)iL_i(f(L_1+3,L_i-1,L_{i+1}+1;A_1+1)-f(L_1+3,L_i-1,L_{i-1}+1;A_1+1),
\end{equation}

\noindent that corresponds to taking into account all the
plaquettes that hit the loop $C$ at links of type $i$, in such a
way that these plaquettes lie outside the area of the loop. The
first term in this expression arises when the links of the
plaquette and the loop that make contact have the same
orientation. In the second term, the links mentioned above have
opposite orientations. We have omitted the dependence of $f$ in
the variables which are not modified by the attachment of the
plaquette.

The second contribution is given by
\begin{equation}
  iL_i(f(L_1+3,L_i-1,L_{i+1}+1;A_i-1,A_{i+1}+1)-f(L_1+3,L_i-1,L_{i-1}+1;A_i-1,A_{i-1}+1)),
\end{equation}
\noindent and is produced by plaquettes that lie on the loop area.

Putting all the contributions together, one arrives to
\begin{eqnarray}
\sum^\infty _{i=0}iL_i\left\{\rule{0ex}{3.5ex} \frac{i}{\beta}f
\right. + f(L_1+3,L_i-1,L_{i+1}+1;A_i-1,A_{i+1}+1)   -
f(L_1+3,L_i-1,L_{i-1}+1;A_i-1,A_{i-1}+1) \nonumber
\\
+(2D-3)(f(L_1+3,L_i-1,L_{i+1}+1;A_1+1) -
f(L_1+3,L_i-1,L_{i-1}+1;A_1+1) \left. \rule{0ex}{3.5ex} \right\}=0
,\label{EcDF}
\end{eqnarray}

\noindent which is a difference equation in two infinite sets of
variables (the $A's$ and the $L's$). Based on the exponential
dependence on area and length that the Wilson Loop should present
in the strong and weak coupling limits, we consider the ansatz
\begin{equation}
  f(L_i;A_i)=\prod^\infty _{i=1}X^{L_i}_i Y^{A_i}_i,
\end{equation}
\noindent which, when substituted into (\ref{EcDF}), produces the
more amenable system of algebraic equations
\begin{equation} \label{EcXY}
  (2D-3)X^3_1X^{-1}_iY_1(X_{i+1}-X_{i-1}) +X^3_1X^{-1}_iY^{-1}_i(X_{i+1}Y_{i+1}-X_{i-1}Y_{i-1})
  +\frac{i}{\beta}=0,\quad i=1,2,\cdots,
\end{equation}
\noindent with $X_0=Y_0=1$. Equation (\ref{EcXY}) can be seen (for
each $i$) as a single equation for two unknown functions: $X_i$
and $Y_i$. Yet, it is not simple enough for our purposes. Hence,
we try separately with length-independent and area-independent
solutions. In the first case, we set $X_i=1\,\,\forall i$, which
yields
\begin{equation}\label{ecuy}
  \beta \{ Y_{i+1}-Y_{i-1} \} +iY_i=0 \quad.
\end{equation}

Taking into account the condition $y_{0}=1$, the solution to this
equation is
\begin{equation}
Y_i(2\beta )=\frac{a \,I_i(2\beta )+b \,(-1)^{i}K_i(2\beta )}{a
\,I_0(2\beta )+b \,K_0(2\beta )},
\end{equation}
where $ I_i$ and $K_i$ are the modified Bessel functions and $a$,
$b$ are arbitrary constants. Since $Y_i(2\beta )$ must be regular
in the strong-coupling limit ($\beta =0$), we must have $b=0$;
hence
\begin{equation} \label{Sol1}
  Y_i=\frac{I_i(2\beta )}{I_0(2\beta )}.
\end{equation}

 On the other hand, the area-independent solution corresponds
to setting $Y_i=1\,\forall i$. This leads to the equation

\begin{equation}
  (D-1)\beta X^3_1(X_{i+1}-X_{i-1})+iX_i=0 \quad .
\end{equation}
 As before, the  admissible solution is
\begin{equation} \label{Sol2}
  X_i=\frac{I_i(4(D-1)\beta X^3_1)}{I_0(4(D-1)\beta X^3_1)} \,\, ,
\end{equation}
that must be understood as follows. For $i=1$, eq. (\ref{Sol2})
gives a transcendental equation for $X_1$. Once this is solved,
eq. (\ref{Sol2}) again allows us to calculate the remaining $X's$.

We recognize that the ``area solution"  (\ref{Sol1}) is the exact
solution of the $D=2$ theory \cite{kogut}, whereas the ``length
solution" (\ref{Sol2}) corresponds to the mean-field solution
(which is dimension-dependent), with $X_1$ being the ``mean link"
\cite{creutz,drou}. In the first case, one obtains a dimension
independent result that, although being the exact solution only
for $D=2$, it is also a  better approximation for the Wilson
functional than the mean-field solution in the strong coupling
regime, even when $D>2$. In fact, the strong-coupling mean-field
solution is $X_1=0$, which corresponds to a vanishing Wilson loop.
Instead, the ``area solution" (\ref{Sol1}) behaves as $<\,W(C)\,>=
(\beta)^{A}$ for $\beta \rightarrow\,0$, which is precisely what
results from strong-coupling expansions in any dimension. On the
other hand, it is well known that the mean-field approximation
provides a qualitatively acceptable description of the weak
coupling region for large enough dimensions, although it
incorrectly predicts phase transitions for $D=2$ and $D=3$.

Summarizing, we have sketched an approach for studying lattice
loop-equations that is based on previous work in the Hamiltonian
formulation. This approach, while being relatively simple, could
capture some relevant features among those that are responsible
for the critical behavior of lattice gauge models. A natural
question at this point is whether or not there is a simple
extension of these ideas to the non-Abelian models, where
loop-equations are considerably more involved. This and other
relevant questions are under work.

This work was supported by DID-USB grant GID-30 and Fonacit grant
G-2001000712.

\end{document}